\documentclass[reqno]{amsart}
\usepackage{graphicx}
\usepackage{epsf}
\usepackage{color}
\def\be{\begin{equation}}
\def\ee{\end{equation}}
\def\ba{\begin{array}}
\def\ea{\end{array}}
\def\bea{\begin{eqnarray}}
\def\eea{\end{eqnarray}}
\def\drm{{\mathrm d}}
\def\erm{{\mathrm e}}

\newcommand{\nc}{\newcommand}
\nc{\tcb}{\textcolor{blue}}
\nc{\tcr}{\textcolor{red}}
\nc{\tcg}{\textcolor{green}}

\nc{\qq}{\qquad\qquad}
\nc{\dis}{\displaystyle}
\nc{\ug}{\; = \;}
\nc{\Ebf}{\mbox{\boldmath $E$}}
\nc{\Bbf}{\mbox{\boldmath $B$}}
\nc{\abf}{\mbox{\boldmath $a$}}
\nc{\vbf}{\mbox{\boldmath $v$}}
\nc{\Fbf}{\mbox{\boldmath $F$}}
\nc{\rbf}{\mbox{\boldmath $r$}}
\nc{\Jbf}{\mbox{\boldmath $J$}}
\nc{\rd}{{\rm d}}
\nc{\dtau}{{\rd\tau}}
\nc{\dt}{{\rd t}}
\nc{\omf}{{\frac{\omega}{\omz}}}
\nc{\tz}{\tau_0}

\begin{document}

\vspace{1truecm}

\title{Majorana and the theoretical problem of photon-electron scattering}

\author{M. Di Mauro}
\address{{\it M. Di Mauro}: Dipartimento di Fisica ``E.R. Caianiello", Universit\'a di Salerno, Via
Giovanni Paolo II, 84084 Fisciano (SA), Italy ({\rm marcodm83@gmail.com})}

\author{S. Esposito}
\address{{\it S. Esposito}: I.N.F.N. Sezione di
Napoli, Complesso Universitario di M. S. Angelo, Via Cinthia,
80126 Naples, Italy ({\rm Salvatore.Esposito@na.infn.it})}%

\author{A. Naddeo}
\address{{\it A. Naddeo}: Dipartimento di Fisica ``E.R. Caianiello", Universit\'a di Salerno, Via
Giovanni Paolo II, 84084 Fisciano (SA), Italy ({\rm adelenaddeo@yahoo.it})}

\begin{abstract}
Relevant contributions by Majorana regarding Compton scattering off free or bound electrons are considered in detail, where a (full quantum) generalization of the Kramers-Heisenberg dispersion formula is derived. The role of intermediate electronic states is appropriately pointed out in recovering the standard Klein-Nishina formula (for free electron scattering) by making recourse to a limpid physical scheme alternative to the (then unknown) Feynman diagram approach. For bound electron scattering, a quantitative description of the broadening of the Compton line was obtained for the first time by introducing a finite mean life for the excited state of the electron system. Finally, a generalization aimed to describe Compton scattering assisted by a non-vanishing applied magnetic field is as well considered, revealing its relevance for present day research.
\end{abstract}

\maketitle



\section{Introduction}

\noindent Among the different phenomena that paved the way to the emergence of the quantum world, the Compton effect certainly played a key role in the acceptance of the photon as the quantum counterpart of an electromagnetic wave \cite{Stuewer1975}. According to Compton's own words \cite{Compton1961}, indeed, the observation of the phenomenon bearing his name just led to ``the concept of X-rays acting as particles'', since the modification of the wavelength of X-rays he observed in his experiments was correctly recognized to be the result of an elastic collision between a photon of given energy with an electron at rest. A standard kinematical analysis of this process, just based on the relativistic energy-momentum conservation, directly led to the Compton formula for the wavelength shift.\footnote{It is probably not well known that, intriguingly enough, as noted by Fermi \cite{teaching}, the exact Compton formula can be deduced also by making recourse to classical (rather than relativistic) kinematics.} 

However, already Compton realized in his experiments that the appearance of an incoherent scattered radiation with a different frequency, in addition to the coherent scattering radiation with the same frequency, was not the only novelty with respect to the classical Thomson scattering of soft X-rays. The measured intensity of the scattered radiation revealed a not at all trivial behavior, the relative importance of the incoherent scattering increasing with the hardness of the X-rays employed and with the corresponding scattering angle: for very hard X-rays impinging on an atomic substance with small atomic weight, at large scattering angles practically only the incoherent radiation is present. The theoretical prediction of the Compton scattering cross-section was, thus, the real problem with which the unborn (and, later, just born) quantum electrodynamics (QED) became involved.

The intensity of the scattering of light waves by a charged particle with mass $m$ and charge $e$ was earlier calculated within Maxwell electrodynamics by Thomson in 1904, obtaining the classical value for the total cross-section:
\begin{equation} \label{thomson}
\sigma = \frac{8 \pi}{3} \, \frac{e^4}{m^2 c^4} \, .
\end{equation}
However, as mentioned above, scattering of hard radiation did not follow this simple relation, and Compton himself in 1923 proposed an {\it ad hoc} formula \cite{Compton1923} -- within a classical picture but with some non-classical ingredients about the frequency shift $\Delta \nu = \nu^\prime$-$\nu$ -- in order to take in some account the experimental observations.
As remarked by C.N. Yang, ``this Compton theory was one of those magic guess works so typical of the 1920's. He knew his theory cannot be entirely correct, so he made the best guess possible'' \cite{Yang}.

The celebrated Klein-Nishina formula \cite{KN}, indeed, came out some years later, after some pioneering works by Kramers and Heisenberg in 1925 \cite{Kramers}, who succeeded in obtaining -- from the correspondence principle -- a dispersion formula for the radiation scattered by atoms, and by Dirac \cite{Dirac1926} \cite{Dirac1927} and, independently, Gordon \cite{Gordon1926}. For the first time they applied quantum mechanics to the Compton problem: the quantized current of the (scalar) electron was calculated (by means of the Schr\"odinger or Klein-Gordon equation) and then used as source of the retarded potential entering the classical expression for the scattering intensity.

Klein and Nishina shared the same strategy of Gordon, but the appearance at the start of 1928 of the Dirac equation allowed the electrons to be described by this relativistic equation, with the obvious consequence that now the electron spin was automatically taken into account. This led to a more complicated formula since, as clearly explained by Compton and Allison \cite{Allison}, the magnetic moment associated with the spin of the electron causes additional scattering, depending both on the photon energy (or frequency) and scattering angle $\theta$. In terms of the differential cross section, i.e. the ratio of the number of scattered photons into the unit solid angle $\Omega$ over the number of incident photons,\footnote{The scattering intensity $I$ of the secondary radiation at a distance $r$ from the emitting electron is directly related to the differential cross section: just multiply ${{\rm d} \sigma}/{{\rm d} \Omega}$ by $r$, divide by the intensity of the incident plane wave and multiply by $\nu^\prime/\nu$.} the result was the following:
\begin{equation} \label{KNformula}
\frac{{\rm d} \sigma}{{\rm d} \Omega} = \frac{1}{2} \, \frac{e^4}{m^2 c^4} \left( \frac{\nu^\prime}{\nu} \right)^2 \left( \frac{\nu^\prime}{\nu} + \frac{\nu}{\nu^\prime} - \sin^2 \theta \right) . 
\end{equation}
This was a remarkable formula, since it was immediately realized that it turned out to be in agreement with the experimental data about the absorption of X-rays by matter \cite{Allison}. Nonetheless, it was obtained still by means of a semiclassical method, where the quantum nature of the electromagnetic field was not at all taken into account. A full quantum approach (quantized radiation and matter fields) appeared soon after, in 1930, when Waller \cite{Waller} and, especially, Tamm \cite{Tamm} rederived the Klein-Nishina formula in a fully consistent approach, by adopting the newly discovered quantum field theory formalism of Heisenberg and Pauli \cite{HP}. 

The main point was that, contrary to Klein and Nishina, the Compton scattering revealed to be a second-order effect, where electron intermediate states are present to bridge from the photon absorption process to that of re-emission of another photon by the electron. This result followed from the application of the time-dependent perturbation scheme of Dirac \cite{Dirac1927b}, the intermediate states being required by the interaction term linear in the electromagnetic field that prevents a direct transition from the initial to the final state. The surprising feature was the {\it necessity} to sum also over negative energy intermediate states of the electron in order to obtain the correct Klein-Nishina formula. Furthermore, Waller paradoxically demonstrated that {\it only} negative energy intermediate states contribute to the classical Thomson scattering formula (\ref{thomson}).

As a result, however, ``the Klein-Nishina formula became the first correct formula of QED discovered by physicists'' \cite{Yang}, despite the peculiarities of the newly born quantum electrodynamics \cite{Weinberg} \cite{Schweber} \cite{Brown} \cite{Pais}, which of course promptly called for a timely experimental validation. Though the Compton effect was not at all the only phenomenon under scrutiny of QED, as explained in \cite{road} the calculation of the related cross-section served nevertheless as a powerful illustration of the attainment of a consistent and also manifestly covariant perturbation calculus of QED processes.

The intensive experimental study, carried out for more than a decade following Compton's initial discovery in 1923, just supported this new theory (and disclosed novel processes, such as pair creation and annihilation \cite{Yang}), but the natural improvement and refinement of the experimental setups also led to the observation of precision effects (in those years, the Compton effect referred to the scattering of a photon on a free electron as well as on bound electrons), which as well called for a theoretical explanation.

In the present paper we unveil the unknown contributions about this subject made by Ettore Majorana \cite{EMER} \cite{EMSEAP} \cite{EMSE} around the same years (end of 1920s), as resulting from the study of his unpublished research notes \cite{volumetti} \cite{quaderni}. The interest in them, indeed, is not only historical in nature but, as common for this author \cite{cup}, pertains also to modern theoretical physics research, given the particularity of the approaches employed, both for the time and for today. This is here illustrated in the treatment of the free electron scattering (Section \ref{free}), as well as in the study of the bound electron scattering (Section \ref{bound}), including also the peculiar case of scattering in presence of an external magnetic field, considered in the known literature only much later than Majorana. Our concluding remarks are reported in the final Section.

\section{Free electron scattering}
\label{free}

\noindent The key idea of the scattering process as a series of successive absorption and emission processes, introduced by Waller and Tamm, is at the basis also of the computations performed by Majorana \cite{quadernifree}. It is not known whether he was aware of the papers of those authors, with whom he shared also the general application of Dirac's theory of dispersion to the radiation scattering problem, but, as a matter of fact, Majorana's calculation approach is quite different from Waller's and Tamm's. In many respects, it is more elegant and closer to modern calculations, though not using the modern formalism of Feynman diagrams (but see below for an intriguing remark), clearly denoting Majorana's mastering of the problem.

As in Waller, the interaction between the quantized electromagnetic radiation and free electrons is described by the Dirac equation\footnote{We choose to adopt the original notation by Majorana, which follows closely that of Dirac, where the usual ``Dirac" matrices are given by $\boldsymbol\alpha=\rho_1\boldsymbol\sigma$, $\beta=\rho_3\mathbb{I}_4$, with $\boldsymbol\sigma$ the Pauli matrices (or, more precisely, 4x4 block matrices with the Pauli matrices on the diagonal), $\mathbb{I}_4$ the 4-dimensional identity matrix, and 
\[
\rho_1=\left(\begin{array}{cc} 0 & \mathbb{I}_2 \\ \mathbb{I}_2 & 0 \\\end{array} \right), \qquad \rho_3=\left(\begin{array}{cc} \mathbb{I}_2 & 0 \\ 0 & -\mathbb{I}_2 \\\end{array}, \right).
\]
Also, the electromagnetic vector potential is denoted with $\mathbf{C}$.}
\bea
\left[\frac{W}{c}+\rho_1\boldsymbol\sigma\cdot\left(\mathbf{p}+\frac{e}{c} \, \mathbf{C}\right) + \rho_3 \, m c\right] \psi =0 \, ,
\eea
which is then the starting point also of Majorana's calculations.

\subsection{Free field quantization}

The electromagnetic field operator is written as:
\bea
\mathbf{C}=\frac{c}{2}\sum_s\sqrt{\frac{2 h}{\pi\nu_s}}\left(a_s\, \mathbf{f}_s + \widetilde{a}_s \, \mathbf{f}_{-s}\right)
\eea
(a tilde denotes hermitian conjugation), where $\widetilde{a}_s=a_{-s}$, $\mathbf{f}_s=\mathbf{f}^*_{-s}$, and the number operator is $\widehat{n}_s=\widetilde{a}_s a_s$, with eigenvalues $n_s$. Field quantization is performed in a cubic box of side $k$, the photon ``wave-functions'' being given by:
\bea
\mathbf{f}_s=\mathbf{k}_s \, \frac{1}{k^{3/2}} \, \erm^{2\pi i \,\boldsymbol\gamma_s\cdot\mathbf{r}/k} \, ,
\eea
where $\boldsymbol\gamma_s$ is a vector with integer components. Similarly, solutions of the free wave equation for the electron in the box are written as:
\bea\label{Wavefunct}
\psi_r=u_r \, \frac{1}{k^{3/2}}\, \erm^{2\pi i\, \boldsymbol\Gamma_r\cdot\mathbf{r}/k} \, ,
\eea
where $u_r$ is a $4-$component normalized Dirac spinor and $\boldsymbol\Gamma$ is another vector of integers. The corresponding electron energies are given by:
\bea
E_r=\pm \,c\,\sqrt{m^2c^2+\frac{h^2}{k^2} \, \Gamma_r^2} \, , \qquad \Gamma_r= \sqrt{\Gamma_{r1}^2 \,+\, \Gamma_{r2}^2 \,+\, \Gamma_{r3}^2} \, .
\eea

\subsection{Interaction hamiltonian}

In order to perform perturbative calculations, the complete Hamiltonian of the system considered by Majorana is splitted in two parts as follows:
\be
\ba{c}
\dis H=H_0 + \mathcal{I},
\\ \\
\dis H_0= -c \rho_1\boldsymbol\sigma\cdot\mathbf{p} - \rho_3 m c^2+ \sum_s \widehat{n}_s h \nu_s \, , \qquad
\mathcal{I}=-e\rho_1\boldsymbol\sigma\cdot\mathbf{C} \, .
\ea
\ee
Notice that, in the free term $H_0$, a quantized electromagnetic field contribution has already been taken into account. 

The generic unperturbed eigenstate is written as $|r;\ldots,n_s,\ldots\rangle$, where $|r\rangle$ is the state vector corresponding to the wave-function (\ref{Wavefunct}), and $n_s$ denotes the number of photons in state $s$. The corresponding unperturbed matrix elements of the free Hamiltonian are simply given by
\bea
\langle\ldots|H_0|\ldots\rangle=E_r+\sum_s n_s h \nu_s 
\eea
(where, however, the zero point energy has been subtracted), while the non-vanishing matrix elements of the interaction term between the unperturbed states are:
\be
\ba{l}
\dis \langle r;\ldots,n_s,\ldots| \, \mathcal{I} \, |r';\ldots,n_s+1,\ldots\rangle \\
\dis \qquad \qquad \qquad \qquad = -\frac{ec}{2k^{3/2}}\,\sqrt{n_s+1}\, \sqrt{\frac{2h}{\pi\nu_s}}\,\, \widetilde{u}_r\rho_1\boldsymbol\sigma\cdot\mathbf{k}_s u_{r'} \, \delta_{\boldsymbol\Gamma_r \! , \, \boldsymbol\Gamma_{r^\prime} +\boldsymbol{\gamma}_s}, \\ \\
\dis \langle r;\ldots,n_s,\ldots|\, \mathcal{I} \, |r';\ldots,n_s-1,\ldots\rangle \\
\dis \qquad \qquad \qquad \qquad = -\frac{ec}{2k^{3/2}}\,\sqrt{n_s}\, \sqrt{\frac{2h}{\pi\nu_s}}\,\, \widetilde{u}_r\rho_1\boldsymbol\sigma\cdot\mathbf{k}_s u_{r'} \, \delta_{\boldsymbol\Gamma_r \! , \,\boldsymbol\Gamma_{r^\prime} -\boldsymbol{\gamma}_s} \, .
\ea
\ee

\subsection{Intermediate states}

The Compton process is the scattering of one photon off one electron, and thus the initial and final states in Majorana's calculations are given by $|1\rangle=|a;1,0\rangle$ and $|2\rangle=|b;0,1\rangle$, respectively, where $a$ and $b$ label the initial and final electron states, while the photon occupation numbers refer just to the $s-$th and $t-$th modes of the quantized radiation field, the other modes being empty. The corresponding energies of these states are, obviously, $E_1=E_a+h\nu_s$ and $E_2=E_b+h\nu_t$. 

Now, it is immediate to see that the matrix element of the perturbation Hamiltonian $\mathcal{I}$ between these two states vanishes, so that no first-order contribution is present, as realized also by Waller and Tamm. The necessity to push the approximation up to second-order terms evidently call for the presence of intermediate states in the matrix element calculations, but here Majorana -- {\it differently} from Waller and Tamm -- realized that only {\it two} possible intermediate states exist that lead to non-vanishing matrix elements of the perturbation Hamiltonian. They refer to $0$- and $2$-photon states, $|i\rangle=|r;0,0\rangle$ and $|i'\rangle=|r';1,1\rangle$, respectively, with corresponding energies $E_i$ and $E_{i'}$.

\begin{figure}
\begin{center}
\includegraphics[width=12cm]{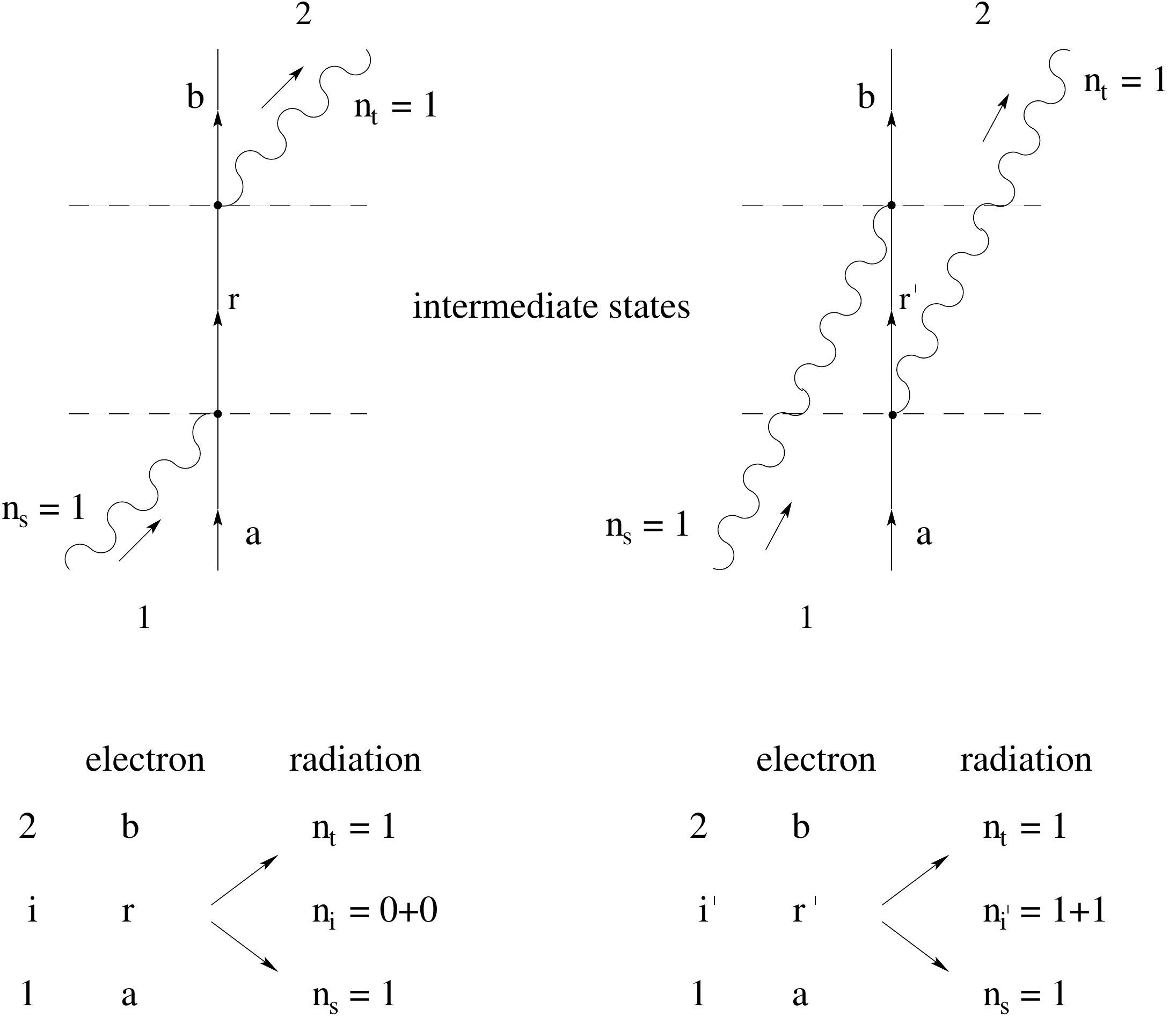}
\caption{Intermediate states in the Compton scattering (at lowest order). Upper panel: standard Feynman diagrams. Lower panel: Majorana's scheme.}
\label{f1}
\end{center}
\end{figure}

In the modern QED formalism \cite{Peskin}, the existence of only two (at second-order) intermediate states corresponds to the fact that only two Feynman diagrams contribute to the Compton scattering (see Fig. \ref{f1}), and it is intriguing that Majorana's plain reasoning behind his straightforward calculations allowed him to recognize this property without the ``visual'' aid now common in quantum field theory textbooks.

\subsection{Dispersion formula}

Majorana then proceeded with his calculations by applying Dirac's time-dependent perturbation theory, just as Waller and Tamm did. The perturbed electron wave-function is expanded in terms of the unperturbed solutions of the wave equation, with time-dependent coefficients $a_i$: at time $t=0$, the initial conditions $a_1=1$, $a_2=a_3=\ldots=0$ hold. For small subsequent times ($t\rightarrow 0$), the transition from the incoming electron to the intermediate states induced by radiation is described by the transition amplitude $a_k$ ($k=i,i^\prime$) solution of the equation:
\bea
\dot{a}_k=-\frac{2\pi\,i}{h}\, \erm^{2\pi\,i(E_k-E_1)t/h}\,I_{k1} \, 
\eea
with $I_{k1}=\langle k| \, \mathcal{I} \, |1\rangle$, that is:
\bea
a_k=-\frac{1}{E_k-E_1}\left(\erm^{2\pi\,i\,(E_k-E_1)t/h}-1\right)I_{k1}.
\eea
One further transition from the given intermediate state to the final state is described by:
\bea
\dot{a}_2=\frac{2\pi i}{h}\sum_k\frac{1}{E_k-E_1}\left( \erm^{2\pi i (E_2-E_1)t/h}-\erm^{2\pi i (E_2-E_k)t/h}\right)I_{2k}I_{k1}\, ,
\eea
whose solution is:
\bea
& & a_2=\sum_k \left[\frac{1}{(E_k-E_1)(E_2-E_1)}\left(\erm^{2\pi\,i\,(E_2-E_1)t/h}-1\right) \nonumber \right.\\ & &  \qquad \qquad \quad\left. - \frac{1}{(E_2-E_k)(E_k-E_1)}\left(\erm^{2\pi\,i\,(E_2-E_k)t/h}-1\right)\right]I_{2k}I_{k1} \, .
\eea
The transition probability from the initial state to the final one is then given by:
\bea \label{prob}
P_{12}=|a_2|^2=\,4\,\frac{\textrm{sin}^2\left[\pi(E_2-E_1)t/h\right]}{(E_2-E_1)^2}\, \left|\sum_k\frac{I_{2k}I_{k1}}{E_k-E_1} \right|^2.
\eea
The pre-factor in the above formula is sharply peaked around $E_2-E_1=0$, so that the dominant contribution to the probability comes from near the resonance $E_1\sim E_2$, obviously ensuring the conservation of energy.\footnote{In the expression above for the transition probability, Majorana takes into account only terms with the resonance denominator equal to $E_2-E_1=0$.} Eq. (\ref{prob}) is, thus, the Majorana (full quantum) form of the Kramers-Heisenberg dispersion formula \cite{Kramers}.

\subsection{Transition probability}

The subsequent calculations of the transition probability for the Compton process make explicit reference to both positive and {\it negative} energy states for the intermediate states $i,i^\prime$:
\be
\ba{l}
\dis E_1=c \, \sqrt{m^2c^2 + \frac{h^2}{k^2}(\boldsymbol\Gamma_a)^2}+h\nu_s\, ,\\ \\
\dis E_2=c \, \sqrt{m^2c^2 + \frac{h^2}{k^2}(\boldsymbol\Gamma_a+\boldsymbol\gamma_s-\boldsymbol\gamma_t)^2}+h\nu_t\, ,\\ \\
\dis E_i=\pm \, c \, \sqrt{m^2c^2 + \frac{h^2}{k^2}(\boldsymbol\Gamma_a+\boldsymbol\gamma_s)^2}\, ,\\ \\
\dis E_{i'}=\pm \, c \, \sqrt{m^2c^2 + \frac{h^2}{k^2}(\boldsymbol\Gamma_a-\boldsymbol\gamma_t)^2}+h\nu_s+h\nu_t
\ea
\ee
(as above, $\nu_s$ and $\nu_t$ refer to the initial and final photon frequencies, respectively). The relevant matrix elements are given by:
\be
\ba{l}
\dis I_{2i}=\langle b;0,1|\, \mathcal{I}\,|r;0,0\rangle=-\frac{ec}{2k^{3/2}} \, \sqrt{\frac{2h}{\pi\nu_t}}\,\,\widetilde{u}_b\rho_1 \boldsymbol\sigma\cdot\mathbf{k}_t u_r\, ,\\ \\
\dis I_{i1}=\langle r;0,0|\, \mathcal{I}\,|a;1,0\rangle=-\frac{ec}{2k^{3/2}}\, \sqrt{\frac{2h}{\pi\nu_s}}\,\,\widetilde{u}_r\rho_1 \boldsymbol\sigma\cdot\mathbf{k}_s u_a\, ,\\ \\
\dis I_{2i'}=\langle b;0,1|\, \mathcal{I}\,|r';1,1\rangle=-\frac{ec}{2k^{3/2}} \, \sqrt{\frac{2h}{\pi\nu_s}}\,\,\widetilde{u}_b\rho_1 \boldsymbol\sigma\cdot\mathbf{k}_s u_{r'} \, ,\\ \\
\dis I_{i'1}=\langle r';1,1|\, \mathcal{I}\,|a;1,0\rangle =-\frac{ec}{2k^{3/2}}\, \sqrt{\frac{2h}{\pi\nu_t}}\,\,\widetilde{u}_{r'}\rho_1 \boldsymbol\sigma\cdot\mathbf{k}_t u_a \, ,
\ea
\ee
where Majorana introduces the following expression for the Dirac spinors:
\be
\ba{ll}
\dis u_a =\left[f_1^a-f_2^a\frac{\boldsymbol\alpha\cdot\mathbf{p}_a}{p_a}\right]u_a^0 \, , \qquad & 
\dis u_b =\left[f_1^b-f_2^b\frac{\boldsymbol\alpha\cdot\mathbf{p}_b}{p_b}\right]u_b^0 \, , \\ \\
\dis u_r =\left[f_1^r\mp f_2^r\frac{\boldsymbol\alpha\cdot\mathbf{p}_r}{p_r}\right]u_r^0 \, , \qquad &
\dis u_{r'}=\left[f_1^{r'}\mp f_2^{r'}\frac{\boldsymbol\alpha\cdot\mathbf{p}_{r'}}{p_{r'}}\right]u_{r'}^0 \, ,
\ea
\ee
with $u_0$ denoting the zero momentum spinors, and the $\mp$ labels the positive/negative energy for the intermediate electron states. The functions $f_1$ and$f_2$ are given by
\bea
f_1^\ell=\sqrt{\frac{1+\sqrt{1+p_\ell^2/m^2c^2}}{2\sqrt{1+p_\ell^2/m^2c^2}}}, \qquad f_2^\ell=\sqrt{\frac{-1+\sqrt{1+p_\ell^2/m^2c^2}}{2\sqrt{1+p_\ell^2/m^2c^2}}}
\eea
($\ell=a,b,r,r'$), and satisfy the relation $|f_1^\ell|^2+|f_2^\ell|^2=1$.

By adopting the standard assumption of the initial electron at rest (i.e. $u_a=u^0_a$, and thus $f_1^a=1$, $f_2^a=0$), after a straightforward but lengthy computation the final transition probability (averaged over initial and summed over final electron spins and photon polarizations) reproduces the standard Klein-Nishina formula (\ref{KNformula}):
\bea
P_{12}\propto \frac{\nu_t}{\nu_s} + \frac{\nu_s}{\nu_t} - \textrm{sin}^2\theta \, , 
\eea
with the initial and final frequencies related (near the resonance) by the Compton formula $\dis \nu_s / \nu_t =1+\frac{h\nu_s}{mc^2}(1-\textrm{cos} \, \theta)$.

\section{Bound electron scattering}
\label{bound}

\noindent Even before the appearance of the Klein-Nishina formula, which settled to a certain extent the problem of the observed scattered intensity in the Compton process, the experimental investigations on it went further and a number of precision effects came out \cite{Stuewer1975}, calling for a thorough physical interpretation and theoretical description. In the decade following the Compton discovery it became clear that X-ray photons interact with atomic electrons substantially in three different ways \cite{Allison}: 1) the photon may be coherently scattered and no change intervenes in the electron state; 2) the photon may be incoherently scattered by the electron, which undergoes a transition to a continuum state; 3) the photon may be incoherently scattered by the electron, which jumps to another bound state. 

The main question emerged, then, that for softer radiation the electron binding had to be taken into account. The basic findings can be summarized as follows. First of all, precision measurements revealed that the Compton line associated to the scattered radiation is broadened \cite{DuMond}, and this was quite promptly associated to the electron motion during the scattering process (Doppler effect due to the electron motion in the atom). Furthermore, the Compton line (for incoherent scattering) resulted to be shifted respect to the Rayleigh one (for coherent scattering) less than expected for free electrons \cite{Ross}, thus revealing a non-vanishing momentum imparted to the atom as the electron is removed. Both these ``kinematical'' effects clearly called for a relaxation of the basic assumptions behind the Klein-Nishina formula or the quantum theori(es) of it. Also, it was detected that the probability of incoherent scattering is decreased at low scattering angles: the energy transfer to a bound atomic electron is suppressed unless the electron gains that amount of energy required for a transition to some available higher energy state. Conversely, the probability of coherent scattering is increased at low angles since, for increasing binding energy, the whole atom absorbs photon momentum, and the probability for coherent Rayleigh scattering increases. It was realized that, for extremely large binding, the Thomson scattering cross section is recovered.

The theoretical analysis of such effects started very early, but remained at a semi-quantitative (or even just qualitative) stage, lacking a general Klein-Nishina-like formula for scattering off bound electrons, which of course depends definitively on the given binding, i.e. on the {\it particular} atom considered.

In 1927, by the use of non-relativistic quantum mechanics, Wentzel \cite{Wentzel1} \cite{Wentzel2} succeeded in obtaining a generalization of the Kramers-Heisenberg dispersion formula to low-energy X-rays and bound electrons (incoherent scattering), showing that the modified line for bound electron scattering is a small continuous spectral distribution ascribed to scattering electrons whose initial state is a discrete (negative energy) level and whose final state is one of the continuum (positive energy) levels. The results he provided applied only to $K$-electrons (as in hydrogen) and, though no explicit formula was given, he deduced that the wavelength modification occasioned by scattering off bound electrons is less than that expected for free electrons.

Wentzel's dispersion formula was corrected, for some peculiarities of incoherent radiation, two years later by Waller and Hartree \cite{WallerHartree}, who performed a quantum mechanical calculation of the intensity of total (coherent plus incoherent) scattering of X-rays by atoms of a mono-atomic gas. The result was that the many-electron problem cannot be obtained as the sum of one-electron problems, since several transitions are forbidden by Pauli principle, some final states being not allowed by it. In the same influential paper of 1930 \cite{Waller} providing the first quantum derivation of the Klein-Nishina formula, Waller considered as well the scattering off bound electrons, but neglecting relativistic and spin effects, and without going further in particular calculations (in a previous paper \cite{Waller1929} he studied the case of electrons bound in a Coulomb potential).

The major general achievement came in 1934, when Bloch \cite{Bloch} relaxed Went\-zel's original assumptions for bound electrons, by describing the motion of the electrons in the atom by hydrogen wave-functions. Further generalization to higher (than $K$) electronic orbits was made by assuming that the forces acting on each electron are approximately described by a Coulomb field with appropriate screening constant. Here the broadening of the Compton line and the defect of the wavelength shift are ascribed to binding forces acting on the scattering electrons. Also, electron interaction with the atomic nucleus not only leads to line broadening, but makes it also asymmetrical, the defect of the Compton shift being shown to be quadratic in the wavelength.

Further subsequent results (up to now) dealt with particular atomic (or molecular) structures, and are of no key interest for our study.

\subsection{Setting the problem}

Majorana did not address all the open questions mentioned above but, quite interestingly, he provided quantitative general results in particular cases, whose physical interpretation has revealed to be long-lasting and particularly illuminating.

The focus was on the scattering of photons on a system of $f$ bound electrons described by the collective wave-function $\psi_a(\mathbf{q}_1,\ldots,\mathbf{q}_f)$ with energy $E_a$ ($a$ labels the corresponding state of the electronic system). The matrix elements of the interaction Hamiltonian between states whose number of photons in a given mode $s$ differs by one are given by:
\be
\ba{l}
\dis \langle a;\ldots,n_s,\ldots| \, \mathcal{I} \, |b;\ldots,n_s+1,\ldots\rangle \\
\dis \qquad \qquad\qquad\qquad =-ec\, \sqrt{\frac{2h(n_s+1)}{2\pi\nu_s}}\int \!\! \widetilde{\psi}_a\sum_{i=1}^f\boldsymbol\alpha^i\cdot\mathbf{f}_s(\mathbf{q}_i) \,\psi_b\, \drm \tau \, , \\ \\
\dis \langle a;\ldots,n_s,\ldots| \, \mathcal{I} \, |b;\ldots,n_s-1,\ldots\rangle \\
\dis \qquad \qquad\qquad\qquad = -ec\, \sqrt{\frac{2h n_s}{2\pi\nu_s}}\int \!\! \widetilde{\psi}_a\sum_{i=1}^f\boldsymbol\alpha^i\cdot\mathbf{f}_s(\mathbf{q}_i)\,
\psi_b\, \drm \tau
\ea
\ee
($\drm \tau$ is the volume element, and $\boldsymbol\alpha^i=\rho_1^i\boldsymbol\sigma^i$). In the following, Majorana adopted the long wavelength (or dipole) approximation for the incident radiation, $\lambda\gg|\mathbf{q}_i|$, for which (again in a cubic box of side $k$)
\bea\label{Approx}
\mathbf{f}_s(\mathbf{q}_i)\sim \mathbf{f}_s(\mathbf{0})=\frac{\mathbf{k_s}}{k^{3/2}}.
\eea

The physical situation he considered was that with the same initial and final energy of the electron system; that is, if the system goes into an excited state, it re-emits exactly the excitation energy. Thus, two possible ways exist for the process to occur, corresponding to two different intermediate states:
\begin{enumerate}
\item  The resonant scattering case occurs, the photon energy being equal to the energy difference between two electron states. The incoming photon is absorbed by the electron system, undergoing a transition into an excited state, and then is re-emitted as the electrons go into their final state. In this case the intermediate state contains no photons and the relevant matrix elements are:
\be\label{MatrixEl1}
\ba{l}
\dis \langle a;0,1|\, \mathcal{I} \, |b;0,0\rangle=-\frac{ec}{k^{3/2}}\, \sqrt{\frac{h}{2\pi\nu_t}}\,\int \!\widetilde{\psi}_a\sum_{i=1}^f \boldsymbol\alpha^i\cdot\mathbf{k}_t \,\psi_b\, \drm \tau \, , \\ \\
\dis \langle b;0,0| \, \mathcal{I} \, |a;1,0\rangle = -\frac{ec}{k^{3/2}}\, \sqrt{\frac{h}{2\pi\nu_s}}\,\int \!\widetilde{\psi}_b\sum_{i=1}^f\boldsymbol\alpha^i\cdot\mathbf{k}_s\,
\psi_a\, \drm \tau \, . 
\ea
\ee
\item Non-resonant scattering occurs: the electron system emits a photon, passing to a lower energy state, and then absorbs another photon, going into its final state. Now the intermediate state contains two photons and the relevant matrix elements are:
\be \label{MatrixEl4}
\ba{l}
\dis \langle a;0,1|\, \mathcal{I} \, |b;1,1\rangle=-\frac{ec}{k^{3/2}}\, \sqrt{\frac{h}{2\pi\nu_s}}\,\int \!\widetilde{\psi}_a\sum_{i=1}^f \boldsymbol\alpha^i\cdot\mathbf{k}_s \,\psi_b\, \drm \tau \, , \\ \\
\dis \langle b;1,1| \, \mathcal{I} \, |a;1,0\rangle=-\frac{ec}{k^{3/2}}\, \sqrt{\frac{h}{2\pi\nu_t}}\,\int \!\widetilde{\psi}_b\sum_{i=1}^f\boldsymbol\alpha^i\cdot\mathbf{k}_t\, \psi_a\, \drm \tau \, . 
\ea
\ee
\end{enumerate}

\subsection{Line broadening}

The transition probability is, then, obtained again with the standard time-dependent perturbation theory, with initial conditions $a_1=1$, $a_2=a_3=\ldots=0$ (at time $t=0$). The transition amplitude to the intermediate states, however, is now solution of the equation:
\bea \label{aboun}
\dot{a}_k=-\frac{2\pi\,i}{h}\,\erm^{2\pi\,i(E_k-E_1)t/h}\,I_{k1} - \frac{1}{2T} \, a_k \, .
\eea
The extra term takes into account the finite mean life of the excited state of the electron system, here denoted with $T$: this approach was introduced by Weisskopf and Wigner in 1930 \cite{Weisskopf} for describing quantitatively general line broadening, and then became traditional.\footnote{For a contemporary reference, see the seminal 1930 lectures of Fermi on the quantum theory of radiation \cite{Fermi}, at Section 8.} However, its application to Compton scattering on bound electrons appears to be due to Majorana, no trace being present in the known literature, as envisaged above. He was probably inspired by his own previous studies about quasi-stationary states in nuclear physics \cite{quasi} \cite{volumetti}, rather than by the subsequent Weisskopf-Wigner paper.

The solution of Eq. (\ref{aboun}) is:
\bea
a_k=-\frac{\erm^{-t/2T}}{E_k-E_1 +\left(h/4\pi i T\right)}\left(\erm^{2\pi\,i\,(E_k-E_1)t/h+t/2T}-1\right)I_{k1} \, ,
\eea
or, for $t\gg T$:
\bea
a_k=-\frac{I_{k1}}{E_k-E_1 +\left(h/4\pi i T\right)} \, \erm^{2\pi\,i\,(E_k-E_1)t/h} \, .
\eea
The amplitude for the transition from the intermediate to the final state is, similarly, the solution of the equation:
\bea
\dot{a}_2=\frac{2\pi i}{h}\sum_k\frac{I_{2k}I_{k1}}{E_k-E_1+\left(h/4\pi i T\right)}\, \erm^{2\pi i (E_2-E_1)t/h} \, ,
\eea
whose solution is:
\bea\label{Solution}
a_2=\left(\sum_k \frac{I_{2k}I_{k1}}{E_k-E_1+\left(h/4\pi i T\right)}\right)\,\frac{\erm^{2\pi\,i\,(E_2-E_1)t/h}-1}{E_2-E_1} \, .
\eea
The transition probability from the initial to the final state is, then, given by:
\bea
P_{12}=|a_2|^2=\,4\,\frac{\textrm{sin}^2\left[\pi(E_2-E_1)t/h\right]}{(E_2-E_1)^2}\, \left|\sum_k\frac{I_{2k}I_{k1}}{E_k-E_1+\left(h/4\pi i T\right)} \right|^2 \, 
\eea
the dominant contribution coming, again, from near the resonance, $E_1\sim E_2$. This formula can, thus, be considered the generalization of Eq. (\ref{prob}) to the case of a non vanishing lifetime $T$ (which further generalizes the original Kramers-Heisenberg formula).

\subsection{Non-vanishing magnetic field}

For some unknown reason, Majorana went on in his analysis by including the effect of a time-varying magnetic field on the bound electrons system. Such a problem was not in the agenda (at the time) for those experimenting on Compton scattering and, then, also for theoretical physicists, so that it is unlikely that Majorana was here stimulated by practical problems. Indeed, any appreciable influence of a magnetic field on the Compton process has some chance to manifest only for very strong magnetic fields, such as  -- in the laboratory case, for sinusoidal fields -- for laser-assisted scattering (for a recent review see \cite{Seipt} and references therein) or, rather, in astrophysical environments (see \cite{Gonthier} and references therein). Even for the simplest case of a constant magnetic field, the only indirect effect is through polarization effects on the electron system \cite{Franz} (see also the more recent paper \cite{Cooper} and references therein) -- that is the magnetic field interacts directly with the electrons, upon which the Compton scattering takes place -- but, again, such phenomenon was considered only later and, in any case, was not the main concern of Majorana's calculations (see below). Instead, Majorana's work seems to have some contact points with another paper of his \cite{variable}, published some years later and related to a different experimental situation studied by his friend and colleague Segr\'e \cite{EMER} \cite{EMSEAP} \cite{EMSE}, or even, alternatively, related to the Raman scattering studied, at the end of 1920's, by Amaldi, Segr\'e and Rasetti (see, for example, \cite{Amaldi29}). However, these are only not well founded speculations, and we do not go further on them, while instead focussing on the results obtained by Majorana, which reveal once more his farsightedness.

The analysis proceeded as above, but the inclusion of a (time-varying) magnetic field $H_x=H_y=0$, $H_z=H(t)$ directed along the positive $z$-axis requires to take into account the magnetic moment contribution $- \mu_z H(t)$ of the electrons in the interaction Hamiltonian. By denoting with $\mu_i =\langle i| \, \mu_z \, |i \rangle$ the (diagonal) matrix elements of the electron magnetic moment, the initial conditions above are replaced by:
\bea
\dot{a}_1=\frac{2\pi i}{h} \, H(t) \, \mu_1 \, a_1 \, , \qquad {\rm or} \qquad a_1=\textrm{exp}\left(\frac{2\pi i}{h}\,\mu_1 \! \! \int \! \! H \drm t\right)
\eea
(and $a_2=a_3=\ldots=0$). The transition amplitude to the intermediate states is, then, the solution of
\bea
\dot{a}_k=-\frac{2\pi\,i}{h}\,I_{k1}\,\erm^{2\pi\,i(E_k-E_1)t/h}\,\erm^{(2\pi i/h)\,\mu_1 \! \int \! H \drm t} - \frac{1}{2T} \, a_k + \frac{2\pi i}{h} \, H \mu_k \, a_k \, ,
\eea
that is:
\bea
& & a_k=\erm^{-t/2T}\erm^{(2\pi i/h) \, \mu_k \! \int \!\! H \drm t}\left(-\frac{2\pi i}{h} \, I_{k1}\right) \nonumber \\
& & \qquad \qquad \times \left[ \int  \! \! \erm^{2\pi i (E_k-E_1)t/h \, + \, t/2T \, + \,  (2\pi i/h) \, (\mu_1-\mu_k) \int \!\! H \drm t} \, \drm t + C \right].
\eea
Therefore, the amplitude for going to the final state is given by:
\be
\dot{a}_2=-\frac{2\pi\,i}{h}\sum_k\,I_{2k}\,\erm^{2\pi\,i(E_2-E_k)t/h} \, a_k  + \frac{2\pi i}{h} \, H \, \mu_2 \, a_2 \, ,
\ee
\be
a_2=-\frac{2\pi i}{h}\erm^{(2\pi i/h) \, \mu_2\int \! \! H \drm t} \sum_k\,I_{2k} \! \int \erm^{2\pi i (E_2-E_k)t/h \, - \, (2\pi i/h) \, \mu_2 \int \!\! H \drm t}\,a_k\,\drm t\, .
\ee
To be definite, Majorana considered the case of a magnetic field oscillating in time with frequency $\nu$, that is 
$ H=H_0 \cos 2\pi\nu t$. In such a case, the ``magnetic term" in the formulas above can be written as follows:
\bea\label{FourierExp}
\erm^{(2\pi i/h) \, \mu \! \int \!\! H \drm t}= \erm^{i A \sin 2\pi\nu t}= \mathcal{I}_0(A) + \mathcal{I}_1(A)e^{2\pi\nu t} + \mathcal{I}_{-1}(A)e^{-2\pi\nu t}+\ldots \, ,
\eea
where $\mathcal{I}_n$ are Bessel functions of the first kind, and $A=H_0\mu/h\nu$. It is apparent from \cite{quaderni} that Majorana limited his analysis to the zeroth order approximation\footnote{We have checked that the first order approximation is very complicated, since it involves a proliferation of additional terms related even to multi-photon intermediate states.} in the Fourier expansion (\ref{FourierExp}), that is an approximately constant magnetic field. For $t\gg T$ we thus have:
\be
a_k^{(0)}=-\frac{I_{k1} \, \mathcal{I}_0(B_k) \, \mathcal{I}_0(A_k)}{E_k-E_1 +\left(h/4\pi i T\right)} \, \erm^{2\pi i (E_k-E_1)t/h} \, ,
\ee
\be
B_k=\frac{H_0 \, \mu_k}{h\nu},\qquad  A_k=\frac{H_0 (\mu_1-\mu_k)}{h\nu} \, , 
\ee
and, finally:
\bea
a_2^{(0)}=\left(\sum_k \frac{I_{2k} \, I_{k1} \, \mathcal{I}_0(B_k) \, \mathcal{I}_0(A_k)}{E_k-E_1+\left(h/4\pi i T\right)}\right) \frac{\mathcal{I}^2_0(C_2) \, (\erm^{2\pi i (E_2-E_1)t/h}-1)}{E_2-E_1} \, ,
\label{eq40}
\eea
with $ C_2={H_0\, \mu_2}/{h\nu}$. This result is the generalization of Eq. (\ref{Solution}) obtained for zero magnetic field, to which obviously reduces in the limit $H_0\rightarrow 0$, since (as noted by Majorana)
\bea
\mathcal{I}_0(x)=1-\frac{x^2}{1\cdot2^2} + \frac{x^4}{2!^2\cdot2^4}+\ldots \, .
\label{exp40}
\eea
Given the approximation of nearly constant fields, that is a very large frequency $\nu$, for small fields $H_0$ the quantities $B_k, A_k, C_2$ in Eq. (\ref{eq40}) assume small values as well, so that the expansion in (\ref{exp40}) evidently shows also that the first non-vanishing magnetic field induced term has the effect to decrease the transition amplitude (and, hence, the scattering intensity) with respect to the zero-field case.

\section{Concluding remarks}

\noindent At the emergence of the quantum description of Nature, quite a relevant role -- though not unique -- was played by the Compton process for the direct detection of photons, i.e. the quanta of the electromagnetic field, as well as for the dynamical description of the effect, which called for a suitable theoretical prediction for the scattering cross section. The experimental validation of the Klein-Nishina formula revealed that quantum mechanics applied successfully also to this electron-photon scattering problem, but the theoretical problem remained of a fully quantum description of the phenomenon, whose solution was the first test bench of the quantum field theory applied to electrodynamics. Indeed, although the QED results obtained by Waller and Tamm just confirmed the semiclassical prediction of the Klein-Nishina formula, the change of perspective was not at all negligible: the Compton scattering resulted to be mediated by electronic intermediate states relating the initial photon absorption process to the final re-emission of another photon by the intervening electron. Also, for the first time, the relevant role of negative energy Dirac states was proved in order to obtain physically meaningful results. Further precision effects revealed by experiments, including scattering off bound -- rather than free -- electrons, also called for a thorough theoretical description but, here, detailed quantitative predictions were generally obscured by mathematical (and physical) technicalities.

In this scenario, quite intriguing emerge the unpublished contributions by Majorana, whose elegant and modern treatment of the basic scattering process reveals very clearly the key physical ideas behind the phenomenon under study. Indeed, if his derivation of the Klein-Nishina formula mathematically resulted just from a full quantum form of the Kramers-Heisenberg dispersion formula he derived, the limpid physical scheme he realized mimicked quite close the later Feynman diagrams approach, based on the existence of only two intermediate electronic states at leading approximation. Major achievements were obtained also for the problem of bound electron scattering, where Majorana was able (for the first time) to quantitatively describe the broadening of the Compton line, again within the now traditional approach by Weisskopf and Wigner, where a finite mean life of the excited state of the electron system is introduced. He was probably led to such an approach by his own pioneering studies about quasi-stationary states in nuclear physics. Finally, and even more intriguing, Majorana unexpectedly studied also the scattering process by bound electrons when a non-vanishing time-varying magnetic field is applied to the system, this phenomenon being considered only in more recent times, when its relevance in some astrophysical environments and in laboratory laser-assisted scattering came out. The general method he devised, though a simple extension of his previous approach, is still completely original, and, even if it was applied just to the simplest case of an approximately constant magnetic field, it reveals a not at all trivial calculation procedure.

It is then easy to predict that what here unveiled will show its relevance quite quickly not only for teaching purposes or in historical studies, but rather also for present day research, where the clean methods developed by Majorana will disclose further interesting results.




\begin{thebibliography}{99}

\bibitem{Stuewer1975}
R.H. Stuewer, The Compton Effect: Turning Point in Physics, Science History Publications, New York, 1975.

\bibitem{Compton1961}
A.H. Compton, Am. J. Phys. {\bf 29} (1961) 817.

\bibitem{teaching}
A. De Gregorio and S. Esposito, Am. J. Phys. {\bf 75} (2007) 781

\bibitem{Stuewer1977}
R.H. Stuewer and M.J. Cooper, {\it Chapter 1}, in B. Williams (ed.), Compton scattering, McGraw-Hill, New York, 1977.

\bibitem{Compton1923}
A.H. Compton, Phys. Rev. {\bf 21} (1923) 483.

\bibitem{Yang}
C.N. Yang, Lect. Notes Phys. {\bf 746} (2008) 393.

\bibitem{KN}
O. Klein and Y. Nishina, Z. Phys. {\bf 52} (1929) 853.

\bibitem{Kramers}
H.A. Kramers and W. Heisenberg, Z. Phys. {\bf 31} (1925) 681.

\bibitem{Dirac1926}
P.A.M. Dirac, Proc. Roy. Soc. {\bf A 111} (1926) 405.

\bibitem{Dirac1927}
P.A.M. Dirac, Proc. Cambridge Philos. Soc. {\bf 23} (1927) 500.


\bibitem{Gordon1926}
W. Gordon, Z. Phys. {\bf 40} (1926) 117.

\bibitem{Allison}
A.H. Compton and S.K. Allison, X-rays in Theory and Experiment, Macmillan, London, 1935.

\bibitem{Waller}
I. Waller, Z. Phys. {\bf 61} (1930) 837.

\bibitem{Tamm}
I. Tamm, Z. Phys. {\bf 62} (1930) 545.

\bibitem{HP}
W. Heisenberg and W. Pauli, Z. Phys. {\bf 56} (1929) 1.

\bibitem{Dirac1927b}
P.A.M. Dirac, Proc. Roy. Soc. {\bf A 114} (1927) 243.

\bibitem{Weinberg}
S. Weinberg, The Quantum Theory of Fields - Foundations (vol. I), Cambridge University Press, Cambridge, 1995.

\bibitem{Schweber}
S.S. Schweber, Q.E.D. and the men who made it, Princeton University Press, Princeton, 1994.

\bibitem{Brown}
L.M. Brown, {\it Introduction: Renormalization 1930-1950}, in L.M. Brown (ed.), Renormalization. From Lorentz to Landau (and beyond), Springer, Berlin, 1993.

\bibitem{Pais}
A. Pais, Inward Bound. Of Matter and Forces in the Physical World, Clarendon Press, Oxford, 1986.

\bibitem{road}
J. Lacki, H. Ruegg and V.L. Telegdi, Stud. Hist. Phil. Sci. Part B {\bf 30} (1999) 457.

\bibitem{EMER}
E. Recami, Il caso Majorana - Epistolario, Documenti, Testimonianze, Di Renzo, Rome, 2011. 

\bibitem{EMSEAP}
S. Esposito, Ann. Phys. (Leipzig) {\bf 17} (2008) 302.

\bibitem{EMSE}
S. Esposito, La cattedra vacante - Ettore Majorana, ingegno e misteri, Liguori, Naples, 2009.

\bibitem{volumetti}
S. Esposito, E. Majorana Jr., A. van der Merwe and E. Recami, Ettore Majorana - Notes on Theoretical Physics, Kluwer, Dordrecht, 2003.

\bibitem{quaderni}
S. Esposito, E. Recami, A. van der Merwe and R. Battiston, Ettore Majorana -- Unpublished Research Notes on Theoretical Physics, Springer, Heidelberg, 2008.

\bibitem{cup}
S. Esposito, The Physics of Ettore Majorana, Cambridge University Press, Cambridge, 2014.

\bibitem{quadernifree}
See Ref. \cite{quaderni}, pp. 104ff.

\bibitem{Peskin}
See, for example, M.E. Peskin and D.V. Schroeder, An Introduction to quantum field theory, Addison-Wesley, Boston, 1995.

\bibitem{DuMond}
J.W.M. DuMond, Rev. Mod. Phys. {\bf 5} (1933) 1.

\bibitem{Ross}
P.A. Ross and P. Kirkpatrick, Phys. Rev. {\bf 46} (1934) 668.

\bibitem{Wentzel1}
G. Wentzel, Z. Phys. {\bf 43} (1927) 1.

\bibitem{Wentzel2}
G. Wentzel, Z. Phys. {\bf 43} (1927) 779.

\bibitem{WallerHartree}
I. Waller and D.R. Hartree, Proc. Roy. Soc. {\bf A 124} (1929) 119.

\bibitem{Waller1929}
I. Waller, Z. Phys. {\bf 58} (1929) 75.

\bibitem{Bloch}
F. Bloch, Phys. Rev. {\bf 46} (1934) 674.

\bibitem{Weisskopf}
V. Weisskopf and E. Wigner, Z. Phys. {\bf 63} (1930) 54.

\bibitem{Fermi}
E. Fermi, Rev. Mod. Phys. {\bf 4} (1932) 87.

\bibitem{quasi}
E. Di Grezia and S. Esposito, Found. Phys. {\bf 38} (2008) 228.

\bibitem{Seipt}
D. Seipt and B. K\"ampfer, Phys. Rev. {\bf A 89} (2014) 023433.

\bibitem{Gonthier}
P.L. Gonthier, A.K. Harding, M.G. Baring, R.M. Costello and C.L. Mercer, Ap. J. {\bf 540} (2000) 907.

\bibitem{Franz}
W. Franz, Ann. Phys. (Berlin) {\bf 425} (1938) 689.

\bibitem{Cooper}
M.J. Cooper, Acta Phys. Pol. {\bf 82} (1992) 137.

\bibitem{variable}
E. Majorana, Nuovo Cim. {\bf 9} (1932) 43.

\bibitem{Amaldi29}
E. Amaldi, Rend. Lincei {\bf 9} (1929) 876.

\end{thebibliography}
\end{document}